\documentclass[draft, preprintnumbers, nofootinbib, preprint, 12pt]{revtex4}
\usepackage{amsmath,amssymb,bm}
\begin{document}
{~}
\vspace{3cm}

\title{Analyticity of Event Horizons of Five-Dimensional Multi-Black Holes with Non-Trivial Asymptotic Structure
\vspace{1cm}
}
\author{
Masashi Kimura\footnote{E-mail:mkimura@sci.osaka-cu.ac.jp}
}
\affiliation{
Department of Mathematics and Physics,  
Graduate School of Science, Osaka City University,
3-3-138 Sugimoto, Sumiyoshi, Osaka 558-8585, Japan
\vspace{3cm}
}

\begin{abstract}
We show that
there exist five-dimensional multi-black hole solutions which have analytic event horizons
when the space-time has non-trivial asymptotic structure,
unlike the case of five-dimensional multi-black hole solutions 
in asymptotically flat space-time.
\end{abstract}

\preprint{OCU-PHYS 298}
\preprint{AP-GR 58}

\pacs{04.50.+h  04.70.Bw}
\date{\today}
\maketitle
{\it Introduction}.---
~~Analyticity of the metric is crucial property
if one wants to investigate
global structure of the space-time
extended across the horizons.
For example, it is well known that
Schwarzschild solution has unique analytic extension across the event horizon.
There are many exact solutions which have analytic horizons~\cite{HE}
\footnote{
In more complicated situation,
some exact solutions have non-analytic horizons~\cite{ChruSin,Brill:1993tm}.
},
and the natural generalizations of four-dimensional single black holes
in asymptotic flat space-time to higher dimensions~\cite{Tangherlini:1963bw,Myers:1986un} 
have also analytic horizons.

In the case of multi-black holes,
four-dimensional Majumdar-Papapetrou(MP) solutions~\cite{Majumdar:1947eu,Papapetrou:1947ib}
have analytic event horizons~\cite{Hartle:1972ya}.
In contrast,
the event horizons of five or higher-dimensional
MP solutions~\cite{Myers:1986rx} are not analytic.
This was first investigated by Gibbons et al~\cite{Gibbons:1994vm} and Welch~\cite{Welch:1995dh},
and recently reinvestigated by Candlish and Reall~\cite{Candlish:2007fh}.
These works would suggest 
event horizons of multi-black holes in higher dimensions are non-analytic in general.
However this is not true 
if the space-time has non-trivial asymptotic structure as discussed later.

In this paper, we focus on five-dimensional space-times whose
topology of boundary of spatial infinity is
not ${\rm S^3}$ but the lens space, ${\rm S^3}/{\mathbb Z}_n$,
and we give simple examples of five-dimensional multi-black hole solutions 
which have analytic event horizons. 

{\it Four-Dimensional Multi-Black Holes in Asymptotically Flat Space-Time}.---
~~First we briefly review that event horizons of the multi-black holes
in four-dimensional asymptotically flat space-time, i.e.,
four-dimensional MP solutions are analytic.
For simplicity, we restrict
ourselves to the MP solutions with two black holes given by
\begin{eqnarray}
ds^2 &=& -H^{-2}dt^2 + H^2 (dr^2 + r^2 d\Omega_{\rm S^2}^2),
\label{4dimMP}
\\
H &=& 1 + \frac{m_1}{r} + \frac{m_2}{\sqrt{r^2 + a^2 - 2 a r \cos \theta}}.
\end{eqnarray}
To remove the divergence of $g_{rr}$ at the event horizon $r=0$, 
we introduce a coordinate across the horizon as
\begin{eqnarray}
dt &=& du + H^{2}dr + W(r,\theta) d\theta,
\label{4dimu}
\\
W(r ,\theta) &=& \int \frac{\partial (H^{2})}{\partial \theta} dr.
\label{4dimW}
\end{eqnarray}
Then the metric (\ref{4dimMP}) becomes
\begin{eqnarray}
ds^2 &=& -H^{-2}du^2 -2 du dr
         -2H^{-2}W du d\theta 
         -2 W dr d\theta
         + W^2 d\theta^2 + H^2 r^2 d\Omega_{\rm S^2}^2.
\end{eqnarray}
In the neighborhood of the horizon $r = 0$, the functions $H$, $H^{-1}$ and $W$ behaves as
\begin{eqnarray}
H &=&\frac{m_1}{r} + O(r^0),
\\
H^{-1}&=& \frac{r}{m_1} + O(r^2),
\\
W &=&
-\frac{2 m_1m_2 \sin \theta}{a^2}r + O(r^2),
\end{eqnarray}
where we choose integral constant of $W$ given by (\ref{4dimW}) as zero.
Then the metric at the horizon becomes
\begin{eqnarray}
ds^2 &=& -2 du dr + m_1^2 d\Omega_{\rm S^2}^2.
\end{eqnarray}
So we can see this coordinate covers the horizon $r=0$.
Moreover, one can easily check all the metric components in this coordinate
are analytic function of $r$ at the horizon $r=0$,
therefore the extension by use of (\ref{4dimu}) and (\ref{4dimW}) is analytic extension across the horizon $r=0$.

{\it Five-dimensional Multi-black holes in asymptotically flat space-time}.---
~~Next,
in the case of five-dimensional MP solutions,
we see that the coordinate across the horizon 
which is introduced like the case of four-dimensional MP solutions
fails to be analytic at the horizon.
The metric of five-dimensional MP solution with two black holes
is given by
\begin{eqnarray}
ds^2 &=& -H^{-2}dt^2 + H (dr^2 + r^2 d\Omega_{\rm S^3}^2),
\\
H &=& 1 + \frac{m_1}{r^2} + \frac{m_2}{r^2 + a^2 - 2 a r \cos \theta}.
\end{eqnarray}
Similar to the equation (\ref{4dimu}),
we introduce a coordinate as
\begin{eqnarray}
dt &=& du + H^{3/2}dr + W(r,\theta) d\theta,
\\
W(r ,\theta) &=& \int \frac{\partial (H^{3/2})}{\partial \theta} dr.
\label{5dimW}
\end{eqnarray}
Then the metric becomes
\begin{eqnarray}
ds^2 &=& -H^{-2}du^2 -2H^{-1/2} du dr
\notag 
\\ && \qquad 
         -2H^{-2}W du d\theta 
         -2H^{-1/2}W dr d\theta
         -H^{-2} W^2 d\theta^2 + H r^2 d\Omega_{\rm S^3}^2.
\end{eqnarray}
In the neighborhood of the horizon $r = 0$, the functions $H$, $H^{-1}$ and $W$ behaves as
\begin{eqnarray}
H &=&\frac{m_1}{r^2} + O(r^0),
\\
H^{-1}&=& \frac{r^2}{m_1} + O(r^4),
\\
W &=&
-\frac{3\sqrt{m_1}m_2 \sin \theta}{a^3}r + O(r^2),
\end{eqnarray}
then all the metric components do not diverge at the horizon.
However, unlike the case of four-dimensional MP solution,
the metric component $g_{ur}$ is zero at $r=0$.
The horizon $r=0$ is still coordinate singularity since 
the metric is degenerate at the horizon in this coordinate.
To remove this coordinate singularity we further introduce the new radial coordinate as
\begin{eqnarray}
\tilde{r} := r^2, \label{tilr}
\end{eqnarray}
then the metric becomes
\begin{eqnarray}
ds^2 &=& -H^{-2}du^2 -H^{-1/2} \tilde{r}^{-1/2} du d\tilde{r}
\notag 
\\ && \qquad 
         -2H^{-2}W du d\theta 
         -H^{-1/2} \tilde{r}^{-1/2} W d\tilde{r} d\theta
         -H^{-2} W^2 d\theta^2 + H \tilde{r} d\Omega_{\rm S^3}^2.
\end{eqnarray}
The behaver of the metric at the horizon is
\begin{eqnarray}
ds^2 
& = &
 -m_1^{-1/2} du d\tilde{r}
         + m_1 d\Omega_{\rm S^3}^2,
\end{eqnarray}
then we can see this coordinate covers the horizon $\tilde{r}=0$.
However, in this coordinate
the function $H$ becomes
\begin{eqnarray}
H = 1 + \frac{m_1}{\tilde{r}} + \frac{m_2}{\tilde{r} + a^2 - 2 a \sqrt{\tilde{r}} \cos \theta},
\label{funcH}
\end{eqnarray}
and we can see the third term in the right hand side of (\ref{funcH})
is not analytic function of ${\tilde{r}}$ at the horizon
$\tilde{r}=0$ because of the existence of $\sqrt{\tilde{r}}$ in the denominator.
This is the essential reason why analyticity is broken in five-dimensional case
unlike the four-dimensional case.
In fact,
from more careful discussions~\cite{Welch:1995dh,Candlish:2007fh},
it is shown that there is no coordinate in which the metric function is analytic at the event horizon.

{\it Five-Dimensional Multi-black holes with non-trivial asymptotic structure}.---
~~Finally, we investigate analyticity of horizons of
multi-black holes constructed on the Gibbons-Hawking space which has non-trivial asymptotic structure, i.e.,
the topology of the spatial infinity is 
lens space ${\rm S}^3/{\mathbb Z}_n$~\cite{Gauntlett:2002nw,Ishihara:2006iv,Ishihara:2006pb}.
The metric with two black holes is given by
\begin{eqnarray}
ds^2 &=&  -H^{-2}dt^2 + H ds^2_{\rm GH},  \label{metfull2}
\\
d{s}^2_{\rm GH}
&=& V^{-1} \left( dr^2 + r^2 d\theta^2 + r^2 \sin^2 \theta d\phi^2 \right)
+V\left(d\zeta + \omega_{\phi} d\phi \right)^2, 
\\
H &=& 1 +\frac{M_1}{r} + \frac{M_2}{\sqrt{r^2 + a^2 - 2a r \cos \theta}},
\\
V^{-1} &=& \epsilon  +\frac{N_1}{r} + \frac{N_2}{\sqrt{r^2 + a^2 - 2a r \cos \theta}},
\\
\omega_{\phi} &=&
 N_1 \cos \theta + \frac{N_2 (-a + r \cos \theta)}{\sqrt{r^2 + a^2  - 2a r \cos \theta}},
\end{eqnarray}
where $ds^2_{\rm GH}$ denotes the metric of the Gibbons-Hawking space~\cite{Gibbons:1979zt}, 
which reduces to the Eguchi-Hanson space for the case $\epsilon = 0$ 
and to the multi-Taub-NUT space for the case $\epsilon =1$
\footnote{
In the special case $\epsilon =1$ and $H = V^{-1}$,
then the metric (\ref{metfull2}) reduces to four-dimensional 
MP solutions with a twisted $S^1$ bundle.
In this case, the horizons are clearly analytic because four-dimensional MP solutions are
analytic on the horizon.}.
Similar to the cases discussed above,
to remove the divergence of $g_{rr}$ at the horizon $r=0$,
we introduce a coordinate as
\begin{eqnarray}
du &=& dt  + H^{3/2}V^{-1/2} dr + W(r,\theta) d\theta,
\\
W(r,\theta) &=&  
\int \frac{\partial (H^{3/2}V^{-1/2})}{\partial \theta} dr.
\label{5dimW2}
\end{eqnarray}
Then the metric becomes
\begin{eqnarray}
ds^2 &=&
        -H^{-2}du^2 -2H^{-1/2}V^{-1/2} du dr 
\notag 
\\ && 
         -2H^{-2}W du d\theta 
         -2H^{-1/2}V^{-1/2} W dr d\theta
         -H^{-2} W^2 d\theta^2 
\notag\\ &&
+r ^2HV^{-1}d\theta^2 
+r ^2 \sin^2\theta HV^{-1}d\phi^2
+HV \bigl(d\zeta+\omega_\phi d\phi\bigr)^2.
\end{eqnarray}
In the neighborhood of the horizon $r = 0$, 
the functions $H$, $H^{-1}$, $V^{-1}$, $V$, $W$ and $\omega_{\phi}$ behave as
\begin{eqnarray}
H &=&\frac{M_1}{r} + O(r^0),
\\
H^{-1}&=& \frac{r}{M_1} + O(r^2),
\\
V^{-1} &=&\frac{N_1}{r} + O(r^0),
\\
V &=& \frac{r}{N_1} + O(r^2),
\\
W &=&
-\sqrt{\frac{M_1}{N_1}}\frac{(3 M_2 N_1 + M_1 N_2) \sin \theta}{2 a^2}r + O(r^2),
\\
\omega_{\phi} &=&
N_1 \cos\theta  - N_2 + O(r),
\end{eqnarray}
then all the metric components do not diverge at the horizon.
Unlike the five-dimensional MP solution,
the metric component $g_{ur}$ is non-zero at the horizon
because of the existence of $V^{-1}$.
Actually, the metric behaves at the horizon as
\begin{eqnarray}
ds^2 
=
-2 \sqrt{\frac{N_1}{M_1}}dudr
+
 M_1 N_1 \left[d\theta^2 + \sin^2\theta d\phi^2 
+ \left(\frac{d\tilde{\zeta}}{N_1} + \cos \theta d\phi \right)^2 \right],
\label{GHhorizon}
\end{eqnarray}
where $\tilde{\zeta} = \zeta -N_2 \phi$,
then this coordinate covers the horizon $r=0$
\footnote{
The spatial cross section of the event horizon in equation (\ref{GHhorizon})
is ${\rm S}^3$ or the lens space ${\rm S}^3/{\mathbb Z}_n$~\cite{Ishihara:2006iv,Ishihara:2006pb}.
}.
Moreover,
the metric is analytic function of $r$ at the event horizon $r=0$ like the four-dimensional MP solution.
Therefore this extension is an analytic extension across the horizon $r=0$.
From this, we can see that
there exist five-dimensional multi-black hole solutions which have analytic event horizons
when the space-time has non-trivial asymptotic structure.
This fact suggests that the analyticity of the event horizon
is tightly related to the asymptotic structure of the space-time.

\section*{Acknowledgements}
The author would like to thank Hideki Ishihara and Ken-ichi Nakao for useful discussions
and careful reading of the manuscript.
This work is supported by a Grant-in-Aid for JSPS Fellows.


\begin{thebibliography}{99}

\bibitem{HE} 
 S.~W.~Hawking and G.~F.~R.~Ellis: {\it The Large Scale
Structure Of Space-Time}, Cambridge University Press,
Cambridge (1973).

\bibitem{Tangherlini:1963bw}
  F.~R.~Tangherlini,
  Nuovo Cim.\  {\bf 27}, 636 (1963).

\bibitem{Myers:1986un}
  R.~C.~Myers and M.~J.~Perry,
  Annals Phys.\  {\bf 172}, 304 (1986).

\bibitem{ChruSin}  P.~T.~Chru\'sciel and D.~B.~Singleton,
    Commun.\ Math.\ Phys.\  {\bf 147}, 137 (1992).

\bibitem{Brill:1993tm}
  D.~R.~Brill, G.~T.~Horowitz, D.~Kastor and J.~H.~Traschen,
  Phys.\ Rev.\  D {\bf 49}, 840 (1994)
  [arXiv:gr-qc/9307014].

\bibitem{Majumdar:1947eu}
  S.~D.~Majumdar,
  Phys.\ Rev.\  {\bf 72}, 390 (1947).

\bibitem{Papapetrou:1947ib}
  A.~Papapetrou,
  Proc.\ Roy.\ Irish Acad.\ (Sect.\ A) A {\bf 51}, 191 (1947).


\bibitem{Hartle:1972ya}
  J.~B.~Hartle and S.~W.~Hawking,
  Commun.\ Math.\ Phys.\  {\bf 26}, 87 (1972).

\bibitem{Myers:1986rx}
  R.~C.~Myers,
  Phys.\ Rev.\  D {\bf 35}, 455 (1987).


\bibitem{Gibbons:1994vm}
  G.~W.~Gibbons, G.~T.~Horowitz and P.~K.~Townsend,
  Class.\ Quant.\ Grav.\  {\bf 12}, 297 (1995)
  [arXiv:hep-th/9410073].

\bibitem{Welch:1995dh}
  D.~L.~Welch,
  Phys.\ Rev.\  D {\bf 52}, 985 (1995)
  [arXiv:hep-th/9502146].

\bibitem{Candlish:2007fh}
  G.~N.~Candlish and H.~S.~Reall,
  Class.\ Quant.\ Grav.\  {\bf 24}, 6025 (2007)
  [arXiv:0707.4420 [gr-qc]].

\bibitem{Gauntlett:2002nw}
  J.~P.~Gauntlett, J.~B.~Gutowski, C.~M.~Hull, S.~Pakis and H.~S.~Reall,
  Class.\ Quant.\ Grav.\  {\bf 20}, 4587 (2003)
  [arXiv:hep-th/0209114].


\bibitem{Ishihara:2006iv}
  H.~Ishihara, M.~Kimura, K.~Matsuno and S.~Tomizawa,
  Class.\ Quant.\ Grav.\  {\bf 23}, 6919 (2006)
  [arXiv:hep-th/0605030].

\bibitem{Ishihara:2006pb}
  H.~Ishihara, M.~Kimura, K.~Matsuno and S.~Tomizawa,
  Phys.\ Rev.\  D {\bf 74}, 047501 (2006)
  [arXiv:hep-th/0607035].

\bibitem{Gibbons:1979zt}
  G.~W.~Gibbons and S.~W.~Hawking,
  Phys.\ Lett.\  B {\bf 78}, 430 (1978).

\end{thebibliography}
\end{document}